\documentclass[seceq]{ptptex}

\title{
Index Theorem in Finite Noncommutative Geometry
}

\author{
Hajime \textsc{AOKI}\footnote{email: haoki@cc.saga-u.ac.jp}
}

\inst{
Department of Physics, Saga University, Saga 840-8502,
Japan 
}

\abst{
Index theorem is formulated in noncommutative geometry with 
finite degrees of freedom by using Ginsparg-Wilson relation.
It is extended to the case where the gauge symmetry is 
spontaneously broken.
Dynamical analysis about topological aspects in gauge theory is also shown.
}

\begin{document}

\maketitle

\section{Introduction}
Noncommutative (NC) geometry is interesting
since it is related to string theories and matrix models
\cite{CDS,NCMM,Seiberg:1999vs},
and it may capture some nature of quantum gravity.
It is also a candidate for a new regularization 
of quantum field theory\cite{Grosse:1995ar}.
Study about topological aspects of gauge theory on it 
is important since
compactification of extra dimensions
with nontrivial index in string theory 
can realize 
chiral gauge theory in our spacetime. 
Ultimately, we hope to realize such a mechanism dynamically,
for instance, in IIB matrix model \cite{IKKT}
where spacetime structure has been studied intensively
\cite{Aoki:1998vn, Nishimura:2001sx}.
Unusual properties of NC geometry may also 
provide a solution of strong CP problem 
and baryon asymmetry of the universe.

Index theorem plays a key role in these studies.
While it can be proved in theories with infinite 
degrees of freedom\cite{Atiyah:1971rm, Kim:2002qm},
it becomes a nontrivial issue in finite cases.
This problem was solved in lattice gauge theory
by using Ginsparg-Wilson (GW) relation
\cite{GinspargWilson,Neuberger,Hasenfratz,Luscher},
and this idea has been successfully extended to 
the NC geometry.
We have provided 
a general prescription to construct 
a GW Dirac operator
with coupling to non-vanishing gauge field backgrounds
on general finite NC geometries\cite{AIN2}.
Owing to the GW relation, an index theorem can be proved 
even for finite NC geometries.
The index takes only integer values by construction, 
and it is shown to become the corresponding 
topological charge 
as the number of degrees of freedom is properly taken to infinity.
While explicit construction has been provided
for the fuzzy 2-sphere\cite{balagovi,AIN2} 
and for the NC torus\cite{Nishimura:2001dq,Iso:2002jc},
it is possible for other cases, which will be reported
in future publication.

As a topologically nontrivial configuration on the fuzzy 2-sphere,
we constructed 't Hooft-Polyakov (TP) monopole configuration
\cite{Balachandran:2003ay,AIN3}.
We further presented a mechanism for  
dynamical generation of a nontrivial index,
by showing that 
the TP monopole configurations 
are stabler than the topologically trivial sector
in the Yang-Mills-Chern-Simons matrix model\cite{AIMN}.
However, in order to obtain non-zero indices
for these configurations,
one needs to introduce a projection operator 
in the definition of the index.
We gave an interpretation for the projection operator,
and extended the index theorem to general configurations
which do not necessarily satisfy the equation of motion 
\cite{Aoki:2006wv}.
Since the $U(2)$ gauge theory on the fuzzy sphere 
is generally broken down to 
$U(1) \times U(1)$ gauge theory through Higgs mechanism, 
this generalization shows that the configuration space
on the fuzzy sphere can be classified 
into the topological sectors. 
We take this subject in section 2.
In section 3 we show dynamical analysis
about topological aspects in gauge theory 
on the NC torus.

\section{Index theorem in spontaneously 
symmetry-broken gauge theory on the fuzzy 2-sphere}

We first give an interpretation for the projection operator. 
TP monopole configuration breaks
the $SU(2)$ gauge symmetry down to $U(1)$, 
and the matter field in the fundamental representation 
contains two components, corresponding to $+1/2$ and $-1/2$ electric charge
of the unbroken $U(1)$ gauge group.
Since these two components cancel the index,
we need to introduce the projection operator 
to prevent the cancellation.

We thus generalize the electric charge operator to
\[
T' = \frac{(A_i)^2-\frac{n^2-1}{4}}{\sqrt{[(A_i)^2-\frac{n^2-1}{4}]^2}}.
\]
This definition is valid for general configurations 
$A_i=L_i + \rho a_i$
unless the denominator has zero-modes.
Here $L_i$ are $n$-dimensional representation of $SU(2)$ algebra
and correspond to coordinates of the fuzzy 2-sphere.
$\rho$ is sphere radius. 
$a_i$ are gauge field in three dimensions, 
and contain gauge field and scalar field 
on the sphere.

We next define modified chirality operators and GW Dirac operator as
\[
\Gamma' = \frac{\{T', \Gamma^R \}}{2} = T' \Gamma^R, \ \ 
\hat\Gamma' = \frac{\{T', \hat\Gamma \}}{\sqrt{\{T', \hat\Gamma \}^2}}, \ \ 
D'_{\rm GW} = -\frac{n}{2} \Gamma' (1 - \Gamma' \hat\Gamma').
\]
For details, please refer to ref. \cite{Aoki:2006wv}.
These operators are weighted by the electric charge operator $T'$, 
which prevent the cancellation of the index.
By the definition, the GW relation
\[
\Gamma' D'_{\rm GW} + D'_{\rm GW} \hat\Gamma' =0
\]
is satisfied, and thus an index theorem
\[
\frac{1}{2} {\rm index}(D'_{\rm GW}) = 
\frac{1}{4} {\cal T}r [\Gamma' + \hat\Gamma']
\]
can be proved.
In the commutative limit, 
where one takes $n$ to infinity and the 
noncommutativity parameter $\alpha$ to zero 
simultaneously
fixing $\rho \sim \alpha n$,
it turns out that
\[
\frac{1}{4} {\cal T}r [\Gamma' + \hat\Gamma']
\to
\frac{\rho^2}{8\pi}\int_{S^2} d\Omega \epsilon_{ijk}
n_i \Bigl( \phi'^a F_{jk}^a 
- \epsilon_{abc} \phi'^a (D_j \phi'^b) (D_k \phi'^c) \Bigr),
\]
where $n_i$ are a unit vector in the normal direction of the sphere,
and $\phi'$ is a normalized scalar field.
This is precisely the topological charge for
the unbroken $U(1)$ gauge symmetry
given by 't Hooft\cite{'tHooft:1974qc}.

This index theorem is valid if the gauge symmetry 
is spontaneously broken to $U(1)$, that is,
when the scalar field
takes non-vanishing vacuum expectation values 
on any points of the sphere.
This is assured by the condition 
$
\bigl[ (A_i)^2-\frac{n^2-1}{4}\bigr]^2
\sim {\cal O}(n^2),
$
which means that all of the eigenvalues 
are of order $n^2$.
Smaller eigenvalues may invalidate the definition of the index, 
while larger eigenvalues may alter the structure of sphere.
Since this condition has both upper and lower bounds,
it gives an extension of the admissibility condition
in the lattice gauge 
theory\cite{Luscher:1981zq}.

Since the formulation is valid for general configurations,
the configuration space can be classified into 
the topological sectors.

\section{Dynamical analysis about topological aspects in 
gauge theory on the NC torus}

It is interesting to study opposite cases where the projection 
operator is not necessary.
It is also important to investigate
dynamics of topological properties in gauge theory
for the aim of the studies mentioned at the beginning in section 1.
We studied these problems
using a lattice formulation of gauge theory on the NC torus
\cite{AMNS} based on
the twisted reduced model\cite{EK,GAO}.
For the 2 dimensional $U(1)$ case,
general classical solutions are known\cite{Griguolo:2003kq}.
We computed the index of the GW Dirac operator
for these classical solutions 
and compared the results with the topological charge
which is obtained as a naive discretization of 
the 1st Chern character\cite{Aoki:2006sb}.  
The two quantities agree when the action is small,
but they take only multiple integer values 
of $N$.
$N$ is the size of the matrices, 
and corresponds to the size of the 2d lattice.
The action for these configurations
is of order $\beta$,
where $\beta \sim 1/g^2$ is the bare coupling constant.
By interpolating the classical solutions, 
we constructed explicit 
configurations
for which the index is of order 1, but
the action becomes of order $\beta N$.

We further performed Monte Carlo simulation 
and obtained the probability distribution of 
the index $\nu$\cite{Aoki:2006zi}.
In the strong coupling region,
the distribution has a form of Gaussian with
a width of order $N$,
which confirms the existence of $\nu \ne 0$ configurations.
Some examples of such configurations are obtained in 
the above interpolation, and reported 
in ref.\cite{Nagao:2005st}
Surprisingly, it turns out that the distribution 
of $\nu$ is asymmetric under $\nu \mapsto -\nu$,
which reflects the parity violation of NC geometry.

In the weak coupling region, however,
the probability for $\nu \ne 0$
decreases rapidly for increasing $N$ and 
for increasing $\beta$.
This is consistent with the above analysis with 
the classical solutions.
In the continuum limit,
we have to send $N$ and $\beta$ to infinity
simultaneously fixing the ratio $\beta / N$,
and thus the distribution approaches to the Kronecker 
delta $\delta_{\nu,0}$.
This result is consistent with the instanton calculation
in the continuum theory\cite{Paniak:2002fi}.
However, it
differs drastically from the results
in the commutative case obtained from 
the lattice simulation\cite{Gattringer:1997bm},
where the distribution is Gaussian with the width of 
the physical extent of the space 
$N/\sqrt{\beta} \sim aN$.
$a$ is lattice spacing.
Intuitively, this situation can be understood as follows.
In the commutative lattice case, 
singular configurations contribute to $\nu \ne 0$ sectors,
which are not realized in the NC geometry
by some smoothing effect.

This property may provide a solution to the strong CP problem,
though some care must be taken when we define the 
$\theta$ vacuum.
As another property of NC geometry,
we found that in general
the probability distribution 
of $\nu$ becomes 
asymmetric under $\nu \mapsto -\nu$. 
One can also twist the boundary condition
to make a topologically nontrivial 
sector dominate in the continuum limit\cite{ANS}.
We expect that these unusual properties of NC geometry
may provide
a dynamical mechanism for realizing chiral fermions in
string theory compactifications, or a mechanism for
generating baryon asymmetry of the universe.
Related works are performed 
using fuzzy spheres 
in the extra dimensions\cite{AIMN,Aschieri:2003vy}.


%

\end{document}